\documentclass[a4paper,11pt,]{article}
\usepackage{amssymb,amsmath}
\usepackage{natbib}
\newtheorem{defin}{Definition}
\newtheorem{defin-theor}{Definition-Theorem}
\newtheorem{defin-rem}{Definition-Remark}
\newtheorem{rem}{Remark}
\newcommand{\rf}[1]{(\ref{#1})}

\newcommand{\bc}{\begin{center}}
\newcommand{\ec}{\end{center}}
\newcommand{\be}{\begin{equation}}
\newcommand{\ee}{\end{equation}}
\newcommand{\bea}{\begin{eqnarray}}
\newcommand{\eea}{\end{eqnarray}}

\newcommand{\bfr}{\begin{flushright}}
\newcommand{\efr}{\end{flushright}}
\newcommand{\bfl}{\begin{flushleft}}
\newcommand{\efl}{\end{flushleft}}
\newcommand{\defeq}{\stackrel{\mbox{\scriptsize def}}{=}}
\newcommand{\comp}{\!\circ\!}
\newtheorem{theor}{Theorem}

\newcommand{\ytp}{\widetilde y}

\begin{document}

\title{General Solution of Functional Equations Defined by\\ Generic Linear-fractional Mappings $F_1: {\bf\rm C}^N\to {\bf\rm C}^N$\\
and by Generic Maps birationally equivalent to $F_1$.}
\author{Konstantin V.~Rerikh\footnote{rerikh@thsun1.jinr.ru }}
\date{}
\maketitle
\begin{center}
{Bogoliubov Laboratory of Theoretical Physics, JINR,
141980, Dubna,\\ The Moscow Region, Russian Federation}
\end{center}
\begin{abstract}
{\rm
We consider a system
 of birational functional equations (BFEs) (or finite-difference equations at $w=m\in{\rm\bf Z}$)
   for  functions $y(w)$ of the form
$$
y(w+1)=F_n(y(w)),\quad y(w):~{\rm\bf C}\to{\rm\bf C}^{N},
 \quad n\defeq {\rm deg}F_n(y), \quad F_n \in {\rm\bf Bir}({\rm\bf C}^{N}),
$$
 where the map $ F_n$ is a given birational one of the
 group of all automorphisms of ${\rm\bf C}^N \to {\rm\bf C}^N$.
The relation of the BFEs with ordinary differential equations is discussed.
  We  present a general solution of the above BFEs
 for $n=1,\quad \forall N$ and of the ones with the map $F_n$ birationally equivalent to $F_1:~~
  F_n\equiv V\comp F_1\comp V^{-1}, \quad \forall V \in {\rm\bf Bir}({\rm\bf C}^{N}).$
}
\end{abstract}



\section{Introduction. Set of the Problem.}

By this paper we start a discussion  of a general problem of integrability
of birational functional equations for
 functions $y(w):~{\rm\bf C}\to{\rm\bf C}^{N} $ in one complex variable $w$
  of the form
\be y(w+1)=F_n(y(w)),\quad y(w):~{\rm\bf C}\to{\rm\bf C}^{N},
 \quad w\in {\rm\bf C}, \quad F_n \in {\rm\bf Bir}({\rm\bf C}^{N}).
\label{eq:bfes1}
 \ee
  For $w=m\in{\rm\bf Z}$ the above BFEs are a dynamical system with a discrete time
  or cascade.
 Here the map
$
  F_n:~y\mapsto y'= F_n(y)=\frac{f_i(y)}{f_{N+1}(y)}, i=(1,2,\ldots,N)$,
$f_i(y)~ \mbox{for}~ \forall~ i$ are polynomials in $y$,
  ${\rm deg}F_n(y)= \max_{i=1}^{N+1} \left\{ {\rm deg}(f_i(y)) \right\}=n,
$
is a given birational one of the
 group of all automorphisms of ${\rm\bf C}^N \to
 {\rm\bf C}^N$ with
 coefficients from ${\rm\bf C}$. By the way, the all said above is valid  and for
 the coefficients from any algebraically closed  field $K$. This fact considerably extends
 the frames of possible applications and using of BFEs \rf{eq:bfes1}.

 The BFEs \rf{eq:bfes1} are
equivalent to BFEs of more general form with a change $(w+1)\to \psi(w)$ where
the map $\psi$ is a given one from the linear-fractional group
 ${\rm\bf Aut}({\rm\bf C})$.
 Actually, the change $w\mapsto \tau(w) =\ln\bigl((w-w_1)/(w-w_2)
\bigr)/\ln(\lambda_1/\lambda_2)$, where the terms $w_i, \lambda_i, i=(1, 2)$
 are the fixed points and the eigenvalues of the map $\psi(w)$ at
 these points
 transform the map $\psi$
 into:
$\quad \psi(\tau):\quad\tau\mapsto \tau'=\tau+1$.

Note also that  discretization  of a standard  autonomous differential equation corresponding
to a vector field \cite{DS-1-88a} ( $\stackrel{.}{x}\defeq dx/d\tau, \tau \in {\bf\rm C}$ )
\be
\stackrel{.}{x}=v(x),\quad x\in U\subset V,\label{eq:vector.field}
\ee
according to $\stackrel{.}{x}\mapsto \frac{ x(\tau+h)-x(\tau)}{h}$ and
$\tau\mapsto hw,\quad x(hw)\mapsto y_h(w)$ gives us a functional equation for $y(w)$:
$$
y_h(w+1)=F_n(y_h(w)),\quad \mbox{where}\quad F_n(y_h(w))\defeq y_h(w) +hv(y_h(w)).
$$
Thus, if we limit ourselves to the vector field $v(y)\defeq \frac{F_n(y)-y}{h}$ where
$F_n(y)\in{\bf Bir(C}^N)$,
then we derive the BFEs \rf{eq:bfes1}.(Of course, there are and other views of discretization.)

The dynamical systems with a discrete time ( at $w=m\in {\bf\rm Z}$) and BFEs
 of the type \rf{eq:bfes1} are an object
 of many investigations of problem of their integrability both
algebraic (see \cite{veselov-89}-\cite{rerikh-2000} and others) and non-algebraic
(see \cite{rerikh-92} -\cite{rerikh-98b}).

The algebraic integrability of BFEs  \rf{eq:bfes1} and  dynamical systems of this type at $N=2$
and for $ \forall n \geq 2$ will be a subject of another paper.

 The problem of integrability of the BFEs \rf{eq:bfes1} for
$n=1$ and for any $N$ is fully solved by the following theorem.

\section{Main Result}
Let us perform a transition from the mapping $F_n$ in $ {\bf\rm C}^N $ to the mapping $\Phi_n$
in ${\bf\rm CP}^N $.

\begin{defin} \label{bir.map}
{\rm Birational maps $\Phi_n, \Phi_n^{-1} $  are the images of the maps
$F_n, F_n^{-1}$ in ${\rm\bf CP}^N$ $y\mapsto z: y_i = z_i/z_{{N+1}},
i\in(1, 2, \cdots, N)$
 and are defined below:
\bea
 \Phi_n:&~& z\mapsto z', \quad z_1':\cdots:z_{{N+1}}'= \phi_1(z):
\cdots: \phi_{{N+1}}(z),\quad z, z'\in {\rm\bf CP}^N,
 \label{eq:fin} \\
\phi_i(z) & = & z_{{N+1}}^nf_i(z_l/z_{{N+1}}), \quad i\in(1, 2, \cdots,
N+1),\quad l\in(1, 2, \cdots, N),
 \eea
 and $\phi_i(z)$ are
 homogeneous polynomials in $z$ without any common factors. The map
 $\Phi_n^{-1}:~z'\mapsto z\sim \phi^{(-1)}(z')$ is defined analogously.
}

\end{defin}

\begin{theor} \label{integ1}
{\rm The system of BFEs \rf{eq:bfes1} at $n={\rm deg}F(y)=1,\quad \phi(z)=Az $, where $A$ is a
complex matrix $(N+1)\times(N+1)$, has a general solution rationally depending on $w$ and
linear-fractionally on $N$ periodic arbitrary functions $I_j(w),\quad j\in(1,\cdots,N)$ of $w$.
We assume that  the matrix $A, {\rm det}(A)\neq 0,$ is preliminary reduced to
 the normal Jordan form (see \cite{gantmacher}):
\be
D=UAU^{-1},\quad D={\rm diag}(D^{(1)},\cdots, D^{(r)}),\quad r\geq 1,\quad
 {\rm dim}(D^{(i)})=k_i, \label{eq:D}
\ee
\be
 D_{s,t}^{(i)}=\lambda_i\delta_{s,t} + \delta_{s,t-1},
 \quad s,t\in(1,\cdots k_i),\quad\sum_{i=1}^{r}k_i=(N+1).\label{eq:Di}
\ee
 Then this solution has the form:
\bea y_i(w) & = & \frac{\sum_{l=1}^{N+1}U_{i,l}^{-1}Y_l(w)}{\sum_{l=1}^{N+1}
U_{N+1,l}^{-1}Y_l(w)}, \quad i=(1,\cdots,N), \label{eq:gs1}\\
Y(w) & = &\{Y^{(1)}_1(w),\cdots,Y^{(1)}_{k_1}(w), \cdots,
Y^{(r)}_1(w),\cdots,Y^{(r)}_{k_r}(w)\}, \nonumber \\
Y_l^{(i)}(w) & = & (\frac{\lambda_i}{\lambda_r})^w\sum_{m=1}^{k_i} C_{l,m}^{(i)}(w)
 I_m^{(i)}(w), \quad \mbox{where} \label{eq:gs2} \\
 I_l^{(i)}(w+1)& = & I_l^{(i)}(w),\quad l=(1,\cdots,k_i),\nonumber\\
C_{l,m}^{(i)} & = & \frac{\Gamma(w+1)\lambda_i^{-(m-l)}}{\Gamma(w+1-m+l) \Gamma(m-l+1)},
\quad \mbox{where} \label{eq:gs3}\\
C_{m,m}^{(i)} & \equiv & 1,\quad C_{l,m}^{(i)}\equiv 0~ \mbox{for}~ l>m.\nonumber
 \eea
 In \rf{eq:gs1}-\rf{eq:gs2} the functions $Y_{k_r}^{r}(w), I_{k_r}^{r}(w)$ are
 identically equal to $1$.
}
\end{theor}

Proof: Let us consider the equation for $z(w),\quad z(w): {\rm\bf C}\to {\rm\bf CP}^N$,
 \be z(w+1)\sim U^{-1}\comp D
 \comp U z(w). \label{eq:gs4}
\ee
 Then supposing $Y(w)={Y^{1}(w), Y^{2}(w),\cdots, Y^{r}(w)}=Uz(w)$
 we have  equations for the function $Y(w): {\bf\rm C}\mapsto{\bf\rm CP}^N$ and
the function $Y^{i}(w):{\bf\rm C}\mapsto{\bf\rm CP}^{k_i}$:
 \be
 Y(w+1)\sim D Y(w), \qquad Y^{i}(w+1)\sim D^{i}Y^{i}(w). \label{eq:gs5}
 \ee
 Remark that the symbol $\sim$ means a projective similarity of
  vectors $z(w+1), Y(w+1), Y^{i}(w+1)$ to  vectors in the right-hand side of equations \rf{eq:gs4},
\rf{eq:gs5}. Then the substitution $Y^{(i)}(w)$ from \rf{eq:gs2}
 transforms equation \rf{eq:gs5} into identity. The functions
 $Y_{k_r}^{(r)}(w)\mbox{ and } I_{k_r}^{(r)}(w)$ are normalized to $1$
  due to a homogeneous
dependence of the numerator and the denominator of expression for $y_i(w)$ \rf{eq:gs1} from the
functions $Y(w)$.
 $\triangleleft$

We can easily generalize this result. Let us introduce the following definition.
\begin{defin} \label{bir.eq.map}
{\rm
Let call birational mapping $F_n$ birationally equivalent to another birational
 map $F_{n'}$ if there exists such a birational mapping $V$ such that the following
equality holds:
\be
F_n=V\comp F_{n'}\comp V^{-1}. \label{eq:gs6}
\ee
}
\end{defin}
$\triangleleft$

The following theorem is valid.

\begin{theor} \label{integn'1}
{\rm
Let BFE \rf{eq:bfes1} be given by the mapping $F_{n}$ birationally equivalent to the mapping
$F_1$ from Theorem \ref{integ1}: $F_n=V\comp F_1\comp V^{-1}$. Then
the general solution of \rf{eq:bfes1} for the function $\ytp(w)$ is equal to
$\ytp(w)=V(y(w))$, where $y(w)$ is given by formulae \rf{eq:gs1}-\rf{eq:gs3}.
}
\end{theor}
 The proof is obvious.

\begin{rem}
{\rm
Let $F_n^k=F_n\comp \cdots\comp F_n$ be a $k$-iteration of the map $F_n$ and $m={\rm deg}V $ be
a degree of the map $V$ from Theorem \ref{integn'1}. Then it is obvious that a boundedness
 of ${\rm deg}F_n^k$ is
a necessary condition  for a birational equivalence of the map $F_n$ to the map $F_1$ since
$F_n^k=V\comp F_1^k\comp V^{-1},\quad{\rm deg}F_1^k =1,$ i.e. ${\rm deg}F_n^k \leq m ^2$.
}
\end{rem}

\section{Acknowledgements}

The author is grateful to R.I.~Bogdanov, V.A.~Iskovskikh, V.V.~Kozlov,
 V.S.~Kulikov, A.N.~Parshin, A.G.~Sergeev, I.R.~Shafarevich, D.V.~Treschev,
I.V.~Volovich and V.S.~Vladimirov for useful discussions and interest in the paper.

\end{document}